\documentclass[preprint,12pt]{elsarticle}

\usepackage{amssymb}
\usepackage{hyperref}
\hypersetup{colorlinks=true,linkcolor=blue,citecolor=red}
\usepackage{amsthm}

\newtheorem{theorem}{Theorem}[section]
\newtheorem{lemma}[theorem]{Lemma}

\theoremstyle{definition}

\theoremstyle{remark}

\newcommand*{\eqref}[1]{(\ref{#1})}

\begin{document}

\begin{frontmatter}



\title{An Explicit Construction of Gauss-Jordan  Elimination Matrix\tnoteref{timu}}

\tnotetext[timu]{This work is partially supported by
NKBRPC-2004CB318003.}


\author{Yi Li\corref{tongxunzuozhe}}
\cortext[tongxunzuozhe]{E-mail address: zm\_liyi@163.com}

\address{Laboratory of Computer Reasoning and Trustworthy
Computing, University of Electronic Science and Technology of China,
Chengdu 610054, China}

\begin{abstract}
\ \ \ \ A constructive approach to get the reduced row echelon form
of a given matrix $A$ is presented. It has been shown that after the
$k$th step of the Gauss-Jordan procedure, each entry $a^k_{ij}(i\neq
j,j>k)$ in the new matrix $A^{k}$ can always be expressed as a ratio
of two determinants whose entries are from the original matrix $A.$
 The new method   also gives a more general
generalization of Cramer's rule than existing methods.
\end{abstract}


\end{frontmatter}


\section{Introduction}
\label{intro}

Gauss-Jordan elimination is a variation of standard Gaussian
elimination in which a matrix is brought to reduced row echelon form
rather merely to triangular form. In contrast to standard Gaussian
elimination, entries above and below the diagonal have to be
annihilated in the process of  Gauss-Jordan elimination. It has been
shown that the  Gauss-Jordan elimination  is considerably less
efficient than Gaussian elimination with backsubstitution when
solving a system of linear equations. Despite its higher cost,
Gauss-Jordan elimination can be preferred in some situations. For
instance, it may be implemented on parallel computers when solving
systems of linear equations \cite{heath}. In addition, it is well
suited for computing the matrix inverse.

 Applying Gauss-Jordan elimination to a given matrix $A,$ we
denote by  $A^{k}$ the new matrix obtained after $k$th step of
Gauss-Jordan elimination. In the present paper,
 we will show that each entry $a^{k}_{i,j}
(i\neq j,j>k)$ in the matrix $A^{k}$ can always be expressed as a
ratio of two determinants whose entries are from the original matrix
$A.$ In 2002, Gong \emph{ et al.} \cite{gae02} first established a
generalized Cramer's rule, which can be applied to a  problem in
decentralized control systems. However, their method is restricted
to deal with a class of particular systems of linear equations. In
\cite{lh07}, Hugo Leiva has presented another generalization of
Cramer's rule, but the given formula is somewhat complicated.
Different from the two methods mentioned above, our approach can
also be used to directly construct one solution of $AX=b.$ From this
point of view, our method can give a generalized Cramer's rule whose
form is completely different from the existing results. We also hope
that it is useful not only as a theoretical tool, but also as a
practical calculation methods in the linear algebra community.

\section{Main results}

\begin{lemma}\cite{howard80}\label{yinli7} If $M$ is a square matrix and
$a,b,c,d$ are scalars, then
 $$\left|M\right|
 \left|
 \begin{array}{ccc} M&U&V\\
R&a&b\\
S&c&d
 \end{array}
  \right|=
 \left|
\begin{array}{lr}
\left|
 \begin{array}{cc} M&U\\
R&a
 \end{array}
  \right|&\left|
 \begin{array}{cc} M&V\\
R&b
 \end{array}
  \right|\\
\left|
 \begin{array}{cc} M&U\\
  S&c
 \end{array}
  \right|&\left|
 \begin{array}{cc} M&V\\
 S&d
 \end{array}
  \right|
\end{array}
 \right|.
$$
\end{lemma}

Before presenting the  main result, we first offer a recursive
description of Bareiss's standard fraction free Gaussian elimination
\cite{Lee95}.
\begin{equation}\label{chap4:12}
a^{(k)}_{i,j}= \left|
\begin{array}{cccc}
a^{0}_{11} & \cdots & a^{0}_{1,k} & a^{0}_{1,j} \\
\vdots & &  \vdots &\vdots \\
a^{0}_{k,1}  &\cdots & a^{0}_{k,k} & a^{0}_{k,j}\\
a^{0}_{i1} & \cdots &a^{0}_{ik},& a^{0}_{ij}
\end{array} \right|,\;\;\; i>k,j>k.
\end{equation}
$$a^{(-1)}_{0,0}=1,a^{(0)}_{i,j}=a_{i,j}$$
$$a^{(k)}_{i,j}=\frac {a^{(k-1)}_{k,k}a^{(k-1)}_{i,j}-a^{(k-1)}_{i,k}a^{(k-1)}_{k,j}}{a^{(k-2)}_{k-1,k-1}}.$$

In what follows, in order to simplify the discussion, we  also
assume that the leading principal minors of a $n\times m$ matrix $A$
are nonzero.

\begin{theorem}\label{dingli8}
Let $A=(a_{ij})$ be a $n\times m$ matrix with entries from an
arbitrary commutative ring and $A^{k}$$(0\le k \le n)$ is defined as
above. Bring $A$ to reduced row echelon form by Gauss-Jordan
elimination. Then after the $k$th elimination step, each entry
$a^{k}_{i,j} (i\neq j,j>k)$ in $A^{k}$ can be expressed as a ratio
of two determinants whose entries are from the original matrix $A.$
\end{theorem}

\begin{proof}~
Consider the following three cases:

$1).$ Case 1: $i>k,j>k.$ We shall  show that
\begin{equation}\label{chap4:13}
a^{k}_{i,j}=\frac {a^{(k)}_{i,j}}{a^{(k-1)}_{k,k}},\; \; \;
(i>k,j>k).
\end{equation}
By $\eqref{chap4:12},$ it is easy to see that the conclusion is
true. To see this, let us use induction on the elimination step $k$
as follows.

(i) When $k=1,$ it is clear that the equality $\eqref{chap4:13}$
holds.

(ii) Now assume that the equality $\eqref{chap4:13}$ is true for
$k.$ Then, when the elimination step is $k+1,$ we have

 $a^{k+1}_{i,j}=\frac
{a^{k}_{k+1,k+1}a^{k}_{i,j}-a^{k}_{i,k+1}a^{k}_{k+1,j}}{a^{k}_{k+1,k+1}}=\frac
{\frac {a^{(k)}_{k+1,k+1}}{a^{(k-1)}_{k,k}} \frac
{a^{(k)}_{i,j}}{a^{(k-1)}_{k,k}}-\frac
{a^{(k)}_{i,k+1}}{a^{(k-1)}_{k,k}} \frac
{a^{(k)}_{k+1,j}}{a^{(k-1)}_{k,k}}}{\frac
{a^{(k)}_{k+1,k+1}}{a^{(k-1)}_{k,k}}}=\frac
{a^{(k+1)}_{i,j}}{a^{(k)}_{k+1,k+1}}.$

This proves the equality \eqref{chap4:13}.

2). Case 2: $i=k,j>k.$ We shall claim that the below formula is
true.
 \begin{equation}\label{chap4:14}
a^{k}_{i,j}=\frac {a^{(k-1)}_{i,j}}{a^{(k-1)}_{k,k}}.
\end{equation}
It is easy to prove this, since we want $a^{k-1}_{k,k}\gets  1,$
according to Gauss-Jordan elimination.

3). Case 3: $i<k,j>k.$ First, Let us construct the following
determinant:
\begin{equation}\label{chap4:15}
a^{(k)}_{i,j}=-\left|
\begin{array}{ccccccc}
a^{0}_{11}&\cdots &a^{0}_{1,i-1}, &a^{0}_{1,i+1} &\cdots &
a^{0}_{1,k},&a^{0}_{1,j}\\
a^{0}_{21}&\cdots &a^{0}_{2,i-1}, &a^{0}_{2,i+1} &\cdots &
a^{0}_{2,k},&a^{0}_{2,j}\\
\vdots & & \vdots &\vdots & &\vdots & \vdots \\
 a^{0}_{k,1}&\cdots &a^{0}_{k,i-1}, &a^{0}_{k,i+1} &\cdots &
a^{0}_{k,k},&a^{0}_{k,j}\\
\end{array}
\right|_{k\times k},\;\;\;i<k,j>k
\end{equation}
Next, we will claim that the following two recursion formulae hold.

Case 3-1. When $i\le k-2,$ we have
\begin{equation}\label{chap4:16}
a^{(k)}_{i,j}=-\frac
{a^{(k-1)}_{k,k}a^{(k-1)}_{i,j}-a^{(k-1)}_{i,k}a^{(k-1)}_{k,j}}{a^{(k-2)}_{k-1,k-1}},\;\;
i\le k-2,j>k.
\end{equation}

Case 3-2. When $i=k-1,$ it follows that
\begin{equation}\label{chap4:17}
a^{(k)}_{i,j}=\frac
{a^{(k-2)}_{k,k}a^{(k-2)}_{i,j}-a^{(k-2)}_{i,k}a^{(k-2)}_{k,j}}{a^{(k-3)}_{k-2,k-2}},\;\;
i=k-1,j>k.
\end{equation}

The proof of the equality $\eqref{chap4:17}:$ Since the row index of
each element in the right-hand side of $\eqref{chap4:17}$ is bigger
than its column index, the formula $\eqref{chap4:12}$ is still
available. By $\eqref{chap4:12},$ we get
$$a^{(k-2)}_{k,k}=\left|
\begin{array}{cccc}
a^{0}_{11}&\cdots&a^{0}_{1,k-2},&a^{0}_{1,k}\\
\vdots& &\vdots &\vdots\\
a^{0}_{k-2,1}&\cdots&a^{0}_{k-2,k-2},&a^{0}_{k-2,k}\\
a^{0}_{k,1}&\cdots&a^{0}_{k,k-2},&a^{0}_{k,k} \end{array}
\right|_{k\times k},\;\; a^{(k-2)}_{k-1,j}=\left|
\begin{array}{cccc}
a^{0}_{11}&\cdots&a^{0}_{1,k-2},&a^{0}_{1,j}\\
\vdots& &\vdots &\vdots\\
a^{0}_{k-2,1}&\cdots&a^{0}_{k-2,k-2},&a^{0}_{k-2,j}\\
a^{0}_{k-1,1}&\cdots&a^{0}_{k-1,k-2},&a^{0}_{k-1,j} \end{array}
\right|_{k\times k}
$$
$$
a^{(k-2)}_{k-1,k}=\left|
\begin{array}{cccc}
a^{0}_{11}&\cdots&a^{0}_{1,k-2},&a^{0}_{1,k}\\
\vdots& &\vdots &\vdots\\
a^{0}_{k-2,1}&\cdots&a^{0}_{k-2,k-2},&a^{0}_{k-2,k}\\
a^{0}_{k-1,1}&\cdots&a^{0}_{k-1,k-2},&a^{0}_{k-1,k} \end{array}
\right|_{k\times k},\;\; a^{(k-2)}_{k,j}=\left|
\begin{array}{cccc}
a^{0}_{11}&\cdots&a^{0}_{1,k-2},&a^{0}_{1,j}\\
\vdots& &\vdots &\vdots\\
a^{0}_{k-2,1}&\cdots&a^{0}_{k-2,k-2},&a^{0}_{k-2,j}\\
a^{0}_{k,1}&\cdots&a^{0}_{k,k-2},&a^{0}_{k,j} \end{array}
\right|_{k\times k}
$$
Partition the above determinants into 4 submatrices respectively, as
follows:
$$M=\left(
\begin{array}{ccc}
a^{0}_{11}&\cdots&a^{0}_{1,k-2}\\
\vdots& &\vdots\\
a^{0}_{k-2,1}&\cdots&a^{0}_{k-2,k-2}
\end{array}
\right) \;\;
a=a^{0}_{k,k},b=a^{0}_{k,j},c=a^{0}_{k-1,k},d=a^{0}_{k-1,j}
$$
$$U=(a^{0}_{1,k},\cdots,a^{0}_{k,k})^{T},\;
V=(a^{0}_{1,j},\cdots,a^{0}_{k-2,j})^{T},
R=(a^{0}_{k,1},\cdots,a^{0}_{k,k})^{T},\;S=(a^{0}_{k-1,1},\cdots,a^{0}_{k-1,k-2})^{T}.$$
In terms of {Lemma}\ref{yinli7},

the right-hand side of $\eqref{chap4:17}$ $=\left|
\begin{array}{ccccc}
a^{0}_{11}&\cdots &a^{0}_{1,k-2},&a^{0}_{1,k},&a^{0}_{1,j}\\
\vdots& &\vdots &\vdots &\vdots\\
a^{0}_{k-2,1}&\cdots &a^{0}_{k-2,k-2},&a^{0}_{k-2,k},&a^{0}_{k-2,j}\\
a^{0}_{k,1}&\cdots&a^{0}_{k,k-2}&a^{0}_{k,k},&a^{0}_{k,j}\\
a^{0}_{k-1,1}&\cdots&a^{0}_{k-1,k-2}&a^{0}_{k-1,k},&a^{0}_{k-1,j}
\end{array}
\right|$ \vspace{0.5cm}

 \hspace{5.0cm}$=a^{(k)}_{k-1,j}.$ \vspace{0.5cm}
\\
\vspace{0.5cm}
 The last equality can be guaranteed by
$\eqref{chap4:15}.$

 A similar but somewhat more complicated method
can be used to establish the proof of $\eqref{chap4:16}.$ According
to $\eqref{chap4:12}$ and $\eqref{chap4:15},$ we have
$$a^{(k-2)}_{k-1,k-1}=\left|
\begin{array}{cccc}
a^{0}_{11}&\cdots&a^{0}_{1,k-2},&a^{0}_{k-1}\\
\vdots & &\vdots&\vdots\\
a^{0}_{k-2,1}&\cdots&a^{0}_{k-2,k-2},&a^{0}_{k-2,k-1}\\
a^{0}_{k-1,1}&\cdots&a^{0}_{k-1,k-2},&a^{0}_{k-1,k-1}
\end{array}
\right|,\;\;
 a^{(k-1)}_{k,k}=\left| \begin{array}{cccc}
a^{0}_{11}&\cdots&a^{0}_{1,k-1},&a^{0}_{1,k}\\
\vdots & &\vdots&\vdots\\
a^{0}_{k-1,1}&\cdots&a^{0}_{k-1,k-1},&a^{0}_{k-1,k}\\
a^{0}_{k,1}&\cdots&a^{0}_{k,k-1},&a^{0}_{k,k}
  \end{array} \right|
$$
$$
a^{(k-1)}_{i,j}=-\left| \begin{array}{ccccccc}
 a^{0}_{11}&\cdots&a^{0}_{1,i-1}&a^{0}_{1,i+1}&\cdots&a^{0}_{1,k-1}&a^{0}_{1,j}\\
 a^{0}_{21}&\cdots&a^{0}_{2,i-1}&a^{0}_{2,i+1}&\cdots&a^{0}_{2,k-1}&a^{0}_{2,j}\\
   \vdots&\vdots &\vdots &\vdots &\vdots &\vdots &\vdots \\
  a^{0}_{k-1,1}&\cdots&a^{0}_{k-1,i-1}&a^{0}_{k-1,i+1}&\cdots&a^{0}_{k-1,k-1}&a^{0}_{k-1,j}
\end{array}
\right|_{(k-1)\times(k-1)}
$$
$$
a^{(k-1)}_{i,k}=- \left|
 \begin{array}{ccccccc}
 a^{0}_{11}&\cdots&a^{0}_{1,i-1}&a^{0}_{1,i+1}&\cdots&a^{0}_{1,k-1}&a^{0}_{1,k}\\
 a^{0}_{21}&\cdots&a^{0}_{2,i-1}&a^{0}_{2,i+1}&\cdots&a^{0}_{2,k-1}&a^{0}_{2,k}\\
   \vdots&\vdots &\vdots &\vdots &\vdots &\vdots &\vdots \\
  a^{0}_{k-1,1}&\cdots&a^{0}_{k-1,i-1}&a^{0}_{k-1,i+1}&\cdots&a^{0}_{k-1,k-1}&a^{0}_{k-1,k}
\end{array}
\right|_{(k-1)\times (k-1)} $$ $$ a^{(k-1)}_{k,j}=\left|
\begin{array}{cccc}
a^{0}_{11}&\cdots&a^{0}_{1,k-1},&a^{0}_{1,j}\\
\vdots& & \vdots &\vdots\\
a^{0}_{k-1,1}&\cdots&a^{0}_{k-1,k-1},&a^{0}_{k-1,j}\\
a^{0}_{k,1}&\cdots&a^{0}_{k,k-1},&a^{0}_{k,j}
\end{array} \right|_{k\times k}.
$$
Afterwards, expand $a^{(k-2)}_{k-1,k-1}$ along the $i$th column, it
follows that
\begin{eqnarray*}
\begin{array}{rl}
a^{(k-2)}_{k-1,k-1}=&(-1)^{i+1}a^{0}_{1,i}M_1+\cdots+(-1)^{k-2+i}a^{0}_{k-2,i}M_{k-2}+(-1)^{k-1+i}a^{0}_{k-1,i}M_{k-1}\\
 =&\sum_{s=1}^{k-1}{(-1)^{s+i}a^{0}_{s,i}M_s}.
\end{array}
\end{eqnarray*}
Here, $M_s$ is a $(k-2)\times (k-2)$ minor of $a^{(k-2)}_{k-1,k-1}.$
\begin{equation} \label{chap4:18}
\begin{array}{rl}
a^{(k-2)}_{k-1,k-1}a^{(k)}_{i,j}&=-\sum_{s=1}^{k-1}
\big[(-1)^{s+i}a^{0}_{s,i}M_s \bar {a}^{(k)}_{i,j}\big]\\
 &=-\sum_{s=1}^{k-1} \big[(-1)^{s+i}a^{0}_{s,i}M_s (-1)^{k-(s+1)}
\left|
\begin{array}{ccc}
\overline {M}_s&U_s&V_s\\
R_s&a_s&b_s\\
S_s&c_s&d_s
\end{array}
\right| \big]
 \end{array}
 \end{equation}

Let $\overline {M_s}$ be a square matrix whose determinant is $M_s.$
Since the minor $M_s$ obtained by expanding $a^{(k-2)}_{k-1,k-1}$
along the $i$th column is exactly a minor of $a^{(k)}_{i,j},$ then
one always can apply elementary row operations to $\bar
{{a}}^{(k)}_{i,j},$ such that the top left corner of $\bar
{a}^{(k)}_{i,j}$ is exactly $\overline {M}_s.$ Here, $\bar
{{a}}^{(k)}_{i,j}=-a^{(k)}_{i,j}.$

According to Lemma \ref{yinli7}, it follows that
$$
\eqref{chap4:18}=(-1)^{k+i}\sum_{s=1}^{k-1}\big(a^{0}_{s,i} \left|
\begin{array}{cc}
\left|
\begin{array}{cc}
\overline {M}_s&U_s\\
R_s&a_s
\end{array}
\right| &\left|
\begin{array}{cc}
\overline {M}_s&V_s\\
R_s&b_s
\end{array}
\right|\\
\left|
\begin{array}{cc}
\overline {M}_s&U_s\\
S_s&c_s
\end{array}
\right|&\left|
\begin{array}{cc}
\overline {M}_s&V_s\\
S_s&d_s
\end{array}
\right|
\end{array}
\right|
 \big)$$
\begin{equation}\label{chap4:19}=
 (-1)^{k+i}\sum_{s=1}^{k-1}\big(a^{0}_{s,i}
 (-1)^{k-s} \left|
\begin{array}{cc}
a^{(k-1)}_{i,k} &a^{(k-1)}_{i,j}\\
\left|
\begin{array}{cc}
\overline {M}_s&U_s\\
S_s&c_s
\end{array}
\right|&\left|
\begin{array}{cc}
\overline {M}_s&V_s\\
S_s&d_s
\end{array}
\right|
\end{array}
\right|
 \big)
\end{equation}

Here, notice that
$$\left |\overline {M}_s \right |=\left |
\begin{array}{ccccccc}
a^{0}_{11}&\cdots
&a^{0}_{1,i-1}&a^{0}_{1,i+1}&\cdots&a^{0}_{1,k-2}&a^{0}_{1,k-1}\\
\vdots& &\vdots&\vdots& &\vdots&\vdots\\
a^{0}_{s-1,1}&\cdots
&a^{0}_{s-1,i-1}&a^{0}_{s-1£¬i+1}&\cdots&a^{0}_{s-1,k-2}&a^{0}_{s-1,k-1}\\
a^{0}_{s+1,1}&\cdots
&a^{0}_{s+1,i-1}&a^{0}_{s+1£¬i+1}&\cdots&a^{0}_{s+1,k-2}&a^{0}_{s+1,k-1}\\
\vdots& &\vdots&\vdots& &\vdots&\vdots\\
a^{0}_{k-1,1}&\cdots
&a^{0}_{k-1,i-1}&a^{0}_{k-1£¬i+1}&\cdots&a^{0}_{k-1,k-2}&a^{0}_{k-1,k-1}\\
\end{array}
\right |_{(k-2)\times (k-2)}$$ $U_s=(a^{0}_{1k}, \cdots,
a^{0}_{s-1,k}, a^{0}_{s+1,k}, \cdots, a^{0}_{k-1,k})^{\tau},$
$V_s=(a^{0}_{1j}, \cdots, a^{0}_{s-1,j}, a^{0}_{s+1,j}, \cdots,
a^{0}_{k-1,j})^{\tau},$ $$S_s=(a^{0}_{s,1}, \cdots, a^{0}_{s,i-1},
a^{0}_{s,i+1}, \cdots, a^{0}_{s,k-2}, a^{0}_{s,k-1})^{\tau},$$
$$a_s=a^{0}_{s,k},
b_s=a^{0}_{s,j},C_s=a^{0}_{k,k},d_s=a^{0}_{k,j}.$$ Clearly,
$$\left | \begin{array}{cc}
\overline {M}_s&U_s\\
R_s&a_s
\end{array}
\right
|=(-1)^{k-(s+1)}(-a^{(k-1)}_{i,k})=(-1)^{k-s}a^{(k-1)}_{i,k},$$
$$\left | \begin{array}{cc}
\overline {M}_s&V_s\\
R_s&b_s
\end{array}
\right
|=(-1)^{k-(s+1)}(-a^{(k-1)}_{i,j})=(-1)^{k-s}a^{(k-1)}_{i,j}.$$ Let
$$\left | \begin{array}{cc}
\overline {M}_s&U_s\\
S_s&c_s
\end{array}
\right |=Q_s,\;\;\;\;\;\;\;\; \left | \begin{array}{cc}
\overline {M}_s&V_s\\
S_s&d_s
\end{array}
\right |=T_s$$ Hence,
$$\eqref{chap4:19}=\sum^{k-1}_{s=1}[(-1)^{i-s+1}a^{0}_{s,i}(a^{(k-1)}_{ij}Q_s-a^{(k-1)}_{i,k}T_s)].$$
Additionally,
\\

$-(a^{(k-1)}_{k,k}a^{(k-1)}_{i,j}-a^{(k-1)}_{i,k}a^{(k-1)}_{k,j})=$

$-\left| \begin{array}{cccccccc}
a^{0}_{11}&\cdots&a^{0}_{1,i-1}&a^{0}_{1,i}a^{(k-1)}_{i,j}&a^{0}_{1,i+1}&\cdots&a^{0}_{1,k-1}&a^{0}_{1,k}\\
\vdots& &\vdots &\vdots&\vdots& &\vdots&\vdots\\
a^{0}_{k-1,1}&\cdots&a^{0}_{k-1,i-1}&a^{0}_{k-1,i}a^{(k-1)}_{i,j}&a^{0}_{k-1,i+1}&\cdots&a^{0}_{k-1,k-1}&a^{0}_{k-1,k}\\
a^{0}_{k,1}&\cdots&a^{0}_{k,i-1}&a^{0}_{k,i}a^{(k-1)}_{i,j}&a^{0}_{k,i+1}&\cdots&a^{0}_{k,k-1}&a^{0}_{k,k}
\end{array}
\right| +$

$\left| \begin{array}{cccccccc}
a^{0}_{11}&\cdots&a^{0}_{1,i-1}&a^{0}_{1,i}a^{(k-1)}_{i,k}&a^{0}_{1,i+1}&\cdots,&a^{0}_{1,k-1}&a^{0}_{1,j}\\
\vdots& &\vdots &\vdots&\vdots& &\vdots&\vdots\\
a^{0}_{k-1,1}&\cdots&a^{0}_{k-1,i-1}&a^{0}_{k-1,i}a^{(k-1)}_{i,k}&a^{0}_{k-1,i+1}&\cdots&a^{0}_{k-1,k-1}&a^{0}_{k-1,j}\\
a^{0}_{k,1}&\cdots&a^{0}_{k,i-1}&a^{0}_{k,i}a^{(k-1)}_{i,k}&a^{0}_{k,i+1}&\cdots&a^{0}_{k,k-1}&a^{0}_{k,j}
\end{array}
\right|.$
\\

And then, expand the above determinants along the $i$th column, we
have
\begin{equation}\label{chap4:20}-(a^{(k-1)}_{k,k}a^{(k-1)}_{i,j}-a^{(k-1)}_{i,k}a^{(k-1)}_{k,j})=
-\sum_{s=1}^{k}\big(a^{0}_{s,i}(-1)^{i+s}\left| \begin{array}{cc}
a^{(k-1)}_{i,j}&a^{(k-1)}_{i,k}\\
A_s&B_s
\end{array}
\right| \big)
\end{equation}

Thereinto, $B_s=\left| \begin{array}{cc} \overline {M}_s&U_s\\
S_s&c_s \end{array}\right|=Q_s,
A_s=\left| \begin{array}{cc} \overline {M}_s&V_s\\
S_s&d_s \end{array}\right|=T_s,$ and $A_s, B_s$ are two minors
obtained by deleting the $s$th row, the $i$th column from the
determinants $a^{(k-1)}_{k,k}a^{(k-1)}_{ij}$, $
a^{(k-1)}_{i,k}a^{(k-1)}_{k,j}$ respectively.

It is important to notice  that when $s=k,$ we have $A_k\equiv
-a^{(k-1)}_{i,k}$ and $B_k\equiv -a^{(k-1)}_{i,j}.$ Therefore, when
$s=k,$ we have
$$\left|
\begin{array}{cc}
a^{(k-1)}_{i,k}&a^{(k-1)}_{i,j}\\
A_k&B_k \end{array}\right|\equiv 0.$$ So,
$\eqref{chap4:19}=\eqref{chap4:20}.$ The equality $\eqref{chap4:16}$
holds clearly.

Now, we consider the third case: when $i<k,j>k,$ the below equality
holds.

\begin{equation}\label{chap4:21}
{a^{k}_{i,j}=} \left\{
\begin{array}{lr}
\frac {a^{(k)}_{i,j}}{a^{(k-1)}_{k,k}}, &i=k-1, j>k,\\
\frac {(-1)^{k-i+1}a^{(k)}_{i,j}}{a^{(k-1)}_{k,k}}, &i\le k-2, j>k
\end{array}
\right.
\end{equation}

Let us induce on $k$ as follows.\\
 (i). When $k=2,3,4,j>k,$ it is very easy to verify that all the following equalities hold.

$$a^{2}_{1,j}=\frac {a^{(2)}_{1,j}}{a^{(1)}_{2,2}},\;\;a^{3}_{1,j}=-\frac {a^{(3)}_{1,j}}{a^{(2)}_{3,3}},\;\;a^{3}_{2,j}=\frac {a^{(3)}_{2,j}}{a^{(2)}_{3,3}},$$
$$a^{4}_{1,j}=\frac {a^{(4)}_{1,j}}{a^{(3)}_{44}},\;\;\;a^{4}_{2,j}=-\frac {a^{(4)}_{2,j}}{a^{(3)}_{44}}\;\;\;
 a^{4}_{3,j}=\frac {a^{(4)}_{3,j}}{a^{(3)}_{44}}.$$
\\
(ii). Suppose that when the elimination step is $k,$
$\eqref{chap4:21}$ still holds. Then, when the elimination step is
$k+1,$ since $i<k+1,$ thus there are tow cases: $i\le k-1$ and
$i=k.$

(ii-1) When $i\le k-2,$ we have \begin{eqnarray*}
\begin{array}{rcl}a^{k+1}_{i,j}&=&\frac
{a^{k}_{k+1,k+1}a^{k}_{i,j}-a^{k}_{i,k+1}a^{k}_{k+1,j}}{a^{k}_{k+1,k+1}}\\
&=&(-1)^{k-i+1}\frac
{a^{(k)}_{k+1,k+1}a^{(k)}_{i,j}-a^{(k)}_{i,k+1}a^{(k)}_{k+1,j}}{a^{(k-1)}_{k,k}a^{(k)}_{k+1,k+1}}\\
&=&(-1)^{k-i}\frac {a^{(k+1)}_{i,j}}{a^{(k)}_{k+1,k+1}}.
\end{array}
\end{eqnarray*}
The lase equality is guaranteed by $\eqref{chap4:16}.$

(ii-2) When $i=k-1,$ we get
\begin{eqnarray*}
\begin{array}{rcl}a^{k+1}_{ij}&=&\frac
{a^{k}_{k+1,k+1}a^{k}_{k-1,j}-a^{k}_{k-1,k+1}a^{k}_{k+1,j}}{a^{k}_{k+1,k+1}}\\
&=&\frac
{a^{(k)}_{k+1,k+1}a^{(k)}_{k-1,j}-a^{(k)}_{k-1,k+1}a^{(k)}_{k+1,j}}{a^{(k-1)}_{kk}a^{(k)}_{k+1,k+1}}\\
&=&-\frac {a^{(k+1)}_{k-1,j}}{a^{(k)}_{k+1,k+1}}.
\end{array}
\end{eqnarray*}

(ii-3) When $i=k,$ it follows that
\begin{eqnarray*}
\begin{array}{rcl}
a^{k+1}_{k,j}&=&\frac
{a^{k}_{k+1,k+1}a^{k}_{k,j}-a^{k}_{k,k+1}a^{k}_{k+1,j}}{a^{k}_{k+1,k+1}}\\
&=&\frac
{a^{(k)}_{k+1,k+1}a^{(k-1)}_{k,j}-a^{(k-1)}_{k,k+1}a^{(k)}_{k+1,j}}{a^{(k-1)}_{k,k}a^{(k)}_{k+1,k+1}}\\
&=&\frac {\frac
{a^{(k-1)}_{k,k}a^{(k-1)}_{k+1,k+1}-a^{(k-1)}_{k+1,k}a^{(k-1)}_{k,k+1}}{a^{(k-2)}_{k-1,k-1}}a^{(k-1)}_{k,j}
-a^{(k-1)}_{k,k+1}\frac
{a^{(k-1)}_{k,k}a^{(k-1)}_{k+1,j}-a^{(k-1)}_{k+1,k}a^{(k-1)}_{k,j}}{a^{(k-2)}_{k-1,k-1}}}
{a^{(k-1)}_{kk}\frac
{a^{(k-1)}_{k,k}a^{(k-1)}_{k+1,k+1}-a^{(k-1)}_{k+1,k}a^{(k-1)}_{k,k+1}}{a^{(k-2)}_{k-1,k-1}}}
=\frac {a^{(k+1)}_{kj}}{a^{(k)}_{k+1,k+1}}.
\end{array}
\end{eqnarray*}
\\
Thus, the equality $\eqref{chap4:21}$ holds. This completes the
proof of  Theorem \ref{dingli8}.

\end{proof}

According to the above results, we know that after $k$th
Gauss-Jordan elimination step, each $a^{k}_{ij} (i\neq j,j>k)$ in
$A^{k}$ can be represented as a ratio of two determinants, as
follows: \vspace{0.5cm}
\begin{equation}\label{total}
{a^{k}_{i,j}=} \left\{
\begin{array}{ll}
\frac {a^{(k)}_{i,j}}{a^{(k-1)}_{k,k}}, &i>k,j>k,\\
\frac {a^{(k-1)}_{i,j}}{a^{(k-1)}_{k,k}}, &i=k,j>k,\\
\frac {a^{(k)}_{i,j}}{a^{(k-1)}_{k,k}}, &i=k-1, j>k,\\
\frac {(-1)^{k-i+1}a^{(k)}_{i,j}}{a^{(k-1)}_{k,k}}, &i\le k-2, j>k.
\end{array}
\right.
\end{equation}

We believe that many results derived by Gauss-Jordan elimination may
be directly reconstructed by $\eqref{total}$. Clearly, by the above
formula one can also easily construct one solution of $AX=b.$ Thus,
this method gives a generalized Cramer's rule.

\section{Conclusions}

As far as we know, the presented approach has not been published.
This new method due to its distinct features can be used in a wide
range of scientific and engineering problems. For example, it
provides a feasible method to solve a system of linear equations
with parametric coefficients by  polynomial interpolation technique
\cite{yang96}. In addition, this method can be  further developed to
give an explicit expression for the elements of the solution of a
constrained linear systems of equations \cite{wq04}. Finally, it can
also be applied to solve some integer programming problems.




\begin{thebibliography}{00}



\bibitem{gae02}Z.M. Gong, M. Aldeen, L. Elsner, A note on a generalized Cramer's rule, Linear Algebra and its Applications. 340 (2002)
253-254

\bibitem{heath}M.T. Heath, Scientific computing: An Introductory
Survey, McGraw-Hill Higher Education, 1997, 47-55.

\bibitem{howard80}E. Howard, Elementary matrix theory, Courier Dover Publications, New York,
1980.

\bibitem{Lee95}H.R. Lee and B. David Saunder, Fraction free gaussian
elimination  for sparse matrices, Journal of Symbolic Computation,
19 (5):393-402, 1995.

\bibitem{lh07}H. Leiva, A generalization of Cramer's rule and applications to generalized linear differential equations,
Revista Notas de Matematica. vol.3(2), No. 258, 2007, pp.81-99

\bibitem{robinson}S.M. Robinson, A short proof of Cramer's rule,
Math. Magazine 43 (1970) 94-95. Reprinted in : S. Montgomery et al.
(Eds.), Selected Papers on Algebra, Math. Assoc. Amer., 1977, pp.
313-314

\bibitem{wq04}G.R. Wang, S.Z. Qiao, Solving constrained matrix equations and Cramer rule, Applied Mathematics and Computation.
159 (2004) 333-340


\bibitem{yang96}L. Yang, J.Z. Zhang, X.R. Hou, Nonlinear algebraic equation system and automated theorem proving, Shanghai Science and
Technology Education Publishing House, Shanghai, 1996, 78-94 (in
Chinese)



\end{thebibliography}
\end{document}